\documentclass[a4paper]{spie}  

\usepackage[a4paper,top=2.54cm,bottom=4.94cm,left=1.93cm,right=1.93cm]{geometry}
\usepackage{amsmath,amsfonts,amssymb}
\usepackage{mathtools}
\usepackage{graphicx}
\usepackage[version=4]{mhchem}
\usepackage{annotate-equations}
\usepackage{listings}
\usepackage[exponent-product=\cdot,separate-uncertainty=true]{siunitx}
\usepackage[colorlinks=true, allcolors=blue]{hyperref}
\usepackage{cleveref}
\crefname{figure}{Fig.}{Figs.}
\crefname{table}{Tab.}{Tabs.}
\crefname{equation}{Eq.}{Eqs.}
\usepackage{xparse}
\ExplSyntaxOn
\NewDocumentCommand{\ccite}{m}
 {
  \seq_set_split:Nnn \l_tmpa_seq { , } { #1 }
  \int_compare:nNnTF { \seq_count:N \l_tmpa_seq } > { 1 }
    { Refs.~\citenum{#1} }
    { Ref.~\citenum{#1} }
 }
\ExplSyntaxOff

\tikzset{annotate equations/text/.style={font=\normalfont}}
\definecolor{lightgray}{RGB}{204, 204, 204}
\definecolor{yellow}{RGB}{230,159,0} 
\definecolor{red}{RGB}{204,121,167} 
\definecolor{green}{RGB}{0,158,115} 
\definecolor{blue}{RGB}{0,114,178} 
\definecolor{orange}{RGB}{213,94,0} 
\renewcommand\fbox{\fcolorbox{lightgray}{white}}
\title{Spectroscopic Performance of Detectors for Athena's WFI: Measurements and Simulation}

\author[a]{J.~Müller-Seidlitz}
\author[a]{R.~Andritschke}
\author[a]{V.~Emberger}
\author[a]{M.~Bonholzer}
\author[a]{G.~Hauser}
\author[b]{P.~Lechner}
\author[a]{A.~Mayr}
\author[a]{J.~Reiffers}
\author[a]{A.~Schweingruber}
\author[c]{W.~Treberspurg}
\affil[a]{Max-Planck-Institute for Extraterrestrial Physics, Gießenbachstraße~1,~85748~Garching~bei~München,~Germany}
\affil[b]{Semiconductor Laboratory of the Max-Planck-Society, Isarauenweg,~85748~Garching~bei~München,~Germany}
\affil[c]{University of Applied Sciences Wiener Neustadt, Johannes-Gutenberg-Straße~3,~2700~Wiener~Neustadt,~Austria}

\authorinfo{Corresponding author:\\\hspace*{2em}J. Müller-Seidlitz: E-mail: johannes.mueller-seidlitz@mpe.mpg.de, Telephone: +49 (0)89 30 000 3509}

\begin{document} 
\maketitle

\begin{abstract}
The depleted p-channel field effect transistor is the chosen sensor type for the Wide Field Imager of the Athena mission. It will be used in two types of cameras. One will enable observations of a field of view of $\SI{40}{\arcminute}\times\SI{40}{\arcminute}$ by using an array of four $512\times512$ pixel sensors in a $2\times2$ configuration. A second, small one is designed to investigate bright, point-like sources with a time resolution of up to \SI{40}{\us}. Sensors of final size, layout, and technology were fabricated, assembled and characterised. Also, first results from the flight production are available and confirm the excellent performance. In order to be able to estimate the future performance of degraded detectors, a simulation was developed that takes into account the non-analytical threshold effects on the basis of measurement results. We present the measurement analysis and the comparison of simulated and measured values as well as first attempts to use the Monte Carlo simulation to predict performance results based on noise measurements.
\end{abstract}


\section{INTRODUCTION}
\label{sec:intro}

The development of the Wide Field Imager (WFI) has officially been started in 2014 after the European Space Agency (ESA) selected Athena (Advanced Telescope for High-ENergy Astrophysics) as one of their large class missions for the Cosmic Vision program to uncover the mysteries of the hot and energetic universe.\cite{nandra13} The telescope consists of two instruments. In addition to WFI, this is the X-ray Integral Field Unit (X-IFU).\cite{barret23} Its key feature is the outstanding spectroscopic resolution of \SI{\le 4}{\eV} while also featuring a good angular resolution.

\begin{figure}[ht]
	\begin{center}
		\begin{tabular}{c}
			\includegraphics[width=12.5cm]{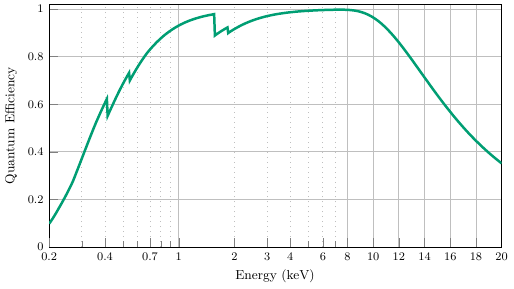}%
		\end{tabular}
	\end{center}
	\caption[qe]
		{ \label{fig:qe} 
		Quantum efficiency of a fully depleted, \SI{450}{\um} thick silicon device with an on-chip filter of \SI{22}{\nm}~\ce{SiO2}, \SI{33}{\nm}~\ce{Si3N4} and \SI{95.15}{\nm}~\ce{Al}. Data are taken from \ccite{henke93}. For a better visualisation of the data, values below \SI{7.4}{\keV} are presented on a logarithmic photon energy scale which covers $3/5$ of the x axis while values above are shown on a linear scale.}
\end{figure}

The Wide Field Imager is an imaging instrument with spectroscopic capability subdivided into two cameras.\cite{mms22} The wide field capability is provided by an array of four detectors---the Large Detector Array. The four individual and independent Large Detectors (LD) comprise $512\times512$ pixels each. The pixel size is \SI{130}{\um}. In total, the Large Detector Array has a field of view of $\SI{40}{\arcminute}\times\SI{40}{\arcminute}$. While a readout time of \SI{\le5}{\ms} for a full frame was originally planned, frame times of \SI{\le2}{\ms} are now feasible with the successful implementation of the drain current readout mode into the readout chain. In the chosen readout scheme of a continuous rolling shutter, the total readout time of one frame depends on the readout time per row. By reducing the number of read-out rows, the exposure time and, therefore, time resolution can be enhanced. This can be done by a window mode readout of a detector, by reducing the detector size or by a parallel readout of rows. To achieve a high time resolution and throughput while preserving a low pile-up for bright sources, this is applied to the concept of the second camera consisting of the Fast Detector (FD). It comprises only $64\times64$ pixels and the readout is split into two halves to improve the frame time even further. It allows for a full frame readout in \SI{\le80}{\us}. By reading only half of both sensor halves, \SI{\le40}{\us} can be achieved. The pixel size of the Fast Detector is the same as for the Large Detectors. The readout time of the latter can be reduced down to \SI{\le250}{\us} in a window mode operation.

The chosen photo sensor is of DEpleted P-channel Field Effect Transistor (DEPFET) type.\cite{kemmer87} It bases on a \SI{450}{\um} fully depleted silicon substrate. It is depleted via the principle of sideward depletion of Silicon Drift Detectors which is also used in special Charge Coupled Devices,\cite{gatti84} so-called pnCCDs, that are used on the X-ray missions XMM-Newton and eRosita. The full depletion of the silicon bulk allows for a high quantum efficiency at \SI{10}{\keV} and above (see \cref{fig:qe}). In addition, the sensor can be illuminated from the back side which allows for a homogeneous and optimised entrance window which is crucial for an optimal spectral response at low photon energies. To reduce the optical loading generated by optical and ultraviolet photons, the back side of the sensor is equipped with an optical blocking filter composed of \SI{20}{\nm}~\ce{SiO2}, \SI{30}{\nm}~\ce{Si3N4} and \SI{86.5}{\nm}~\ce{Al}. To account for inaccuracies in the layer thicknesses, \SI{10}{\percent} were added to each layer to calculate the quantum efficiency for photons in the interval from \SI{0.2}{\keV} to \SI{20}{\keV} in \cref{fig:qe}.

Incident X-ray photons that are absorbed in the silicon crystal lattice of the sensitive detector volume generate electron-hole pairs according to Fano statistics (see \cref{eq:fwhm}).\cite{fano47} The electrons and holes are separated by an externally applied voltage. Holes are drained via the back side contact while electrons are collected at the front side. Each pixel of the active pixel sensor contains a DEPFET readout node. Below the transistor's (external) gate an n-implant forms a potential minimum for electrons. These electrons generate a proportional number of mirror charges in the transistor channel which modifies its conductance. Due to its modelling effect on the transistor behaviour, it is called internal gate. The relation between the change in source-drain-current caused by a change in the signal is given by the charge gain $g_\text{q}$.\cite{lutz99}

\vspace{2em}
\begin{equation}
\label{eq:gq}
	\eqnmark[black]{gq}{g_\text{q}} = \frac{\partial \eqnmark[orange]{ids}{I_\text{DS}}}{\partial \eqnmark[blue]{qsig}{Q_\text{sig}}} \propto -\frac{1}{\eqnmark[green]{gl}{L}^2} \eqnmarkbox[yellow]{muh}{\mu_\text{h}} \eqnmark[red]{uds}{U_\text{DS}}
\end{equation}
\annotate[yshift=-1em,xshift=-1em]{below,left}{gq}{Charge gain}
\annotate[yshift=0.7em]{above,left}{ids}{Source-drain current}
\annotate[yshift=-2em,xshift=-1em]{below,left}{qsig}{Signal charge}
\annotate[yshift=-2em,xshift=1em]{below,right}{gl}{Gate length}
\annotate[yshift=1.5em]{above,right}{muh}{Hole mobility}
\annotate[yshift=-1em,xshift=1em]{below,right}{uds}{Source-drain voltage}
\vspace{3.5em}

\begin{figure}[ht]
	\begin{center}
		\begin{tabular}{c} 
			\fbox{\includegraphics[width=7cm]{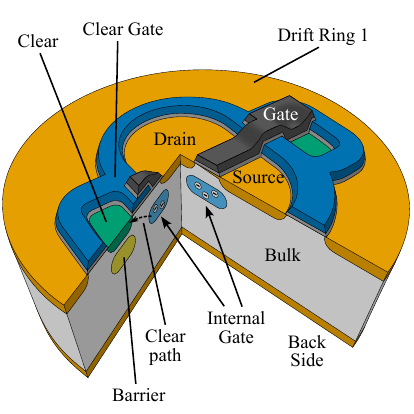}}\hspace{0.5cm}%
			\fbox{\includegraphics[width=7cm]{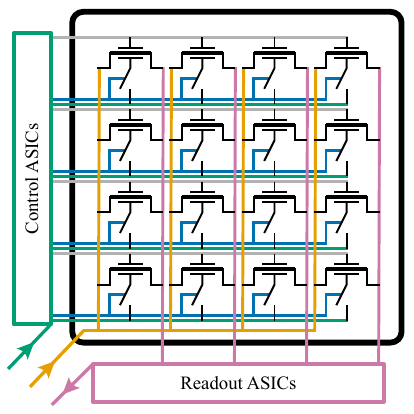}}%
			\\[\bigskipamount]%
			\fbox{\includegraphics[width=7cm]{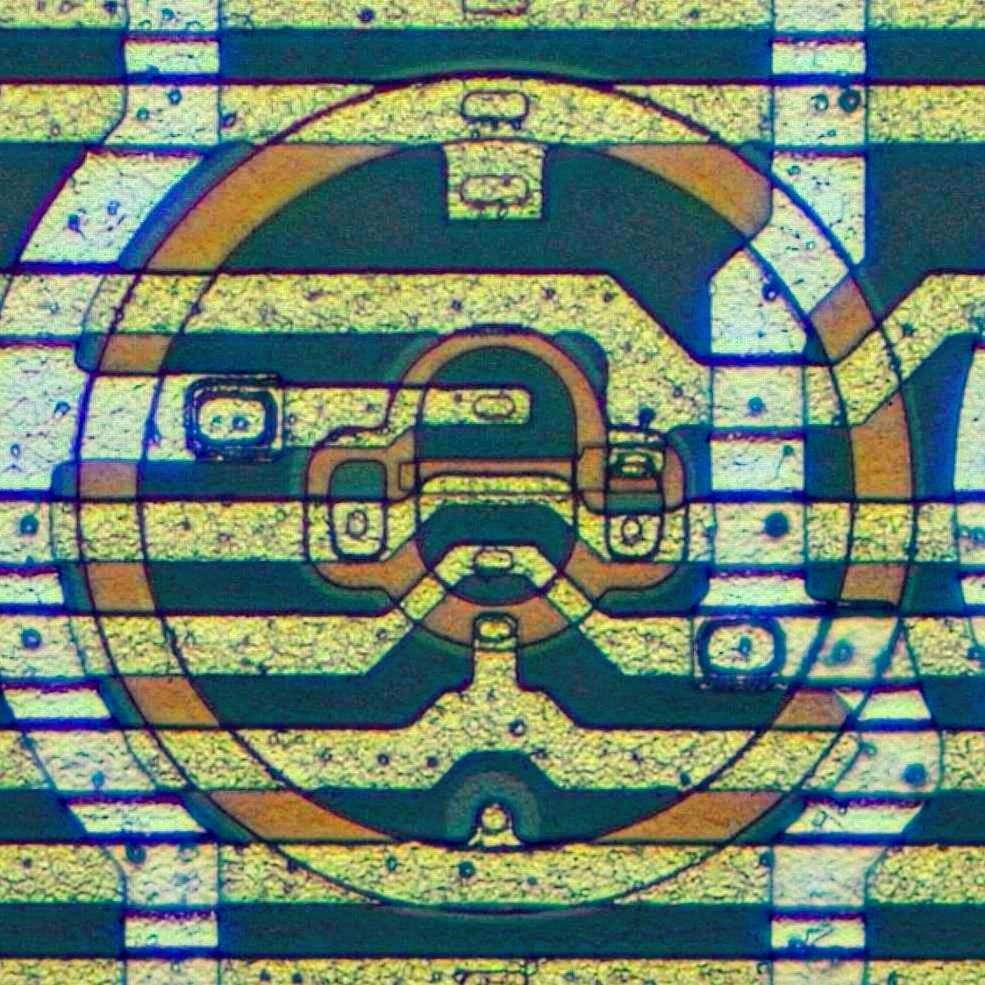}}\hspace{0.5cm}%
			\fbox{\includegraphics[width=7cm]{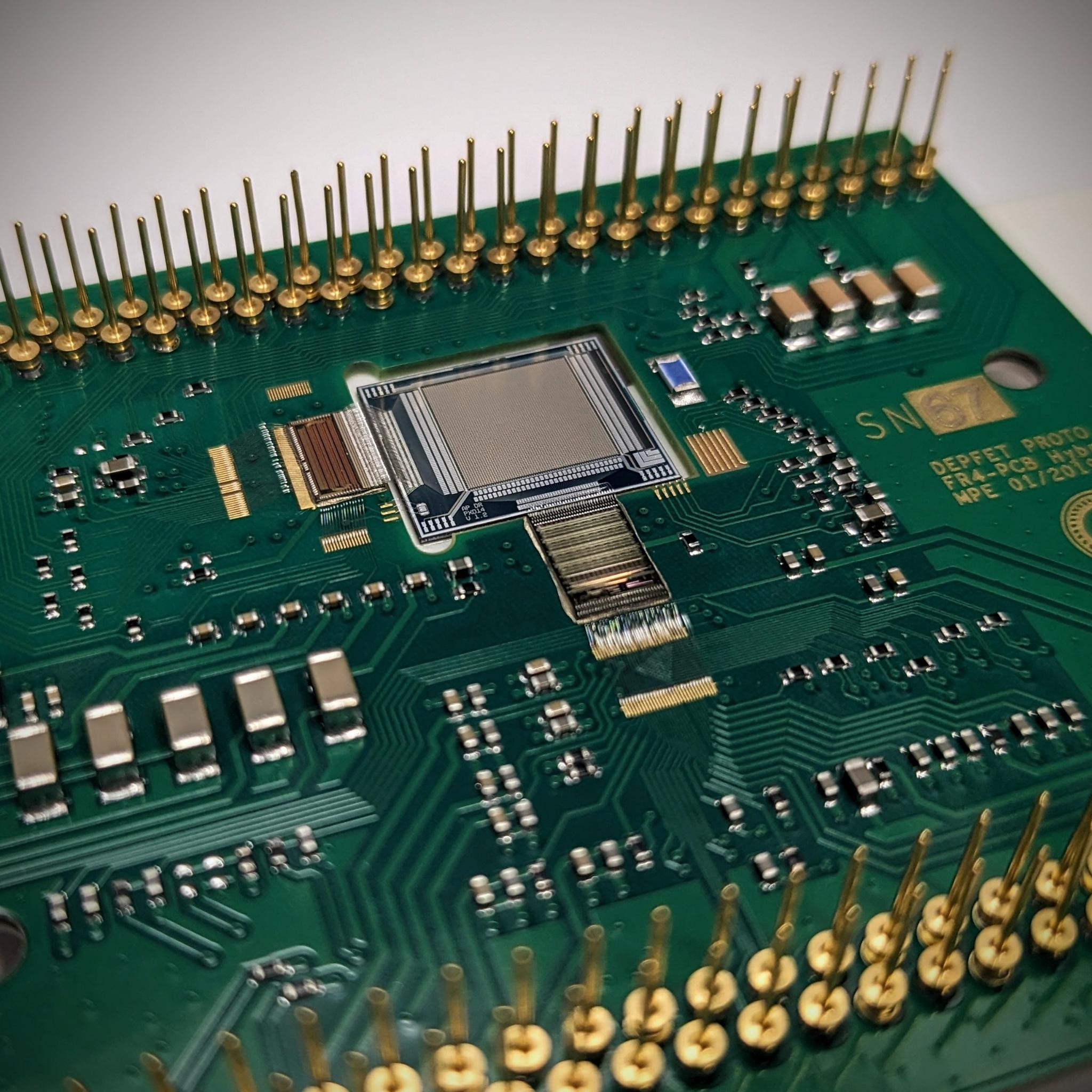}}%
		\end{tabular}
	\end{center}
	\caption[depfet]
		{ \label{fig:depfet} 
		\textit{Upper left}: Layout of a DEPFET as used for the WFI. \textit{Upper right}: Configuration of a detector prepared for rolling shutter readout. Apart from the drain readout nodes, the colour code of the metal connection lines is the same as for the contacts in the DEPFET sketch. \textit{Lower left}: Photo of a single pixel from the flight production. \textit{Lower right}: Prototype detector module with a $64\times64$ pixel DEPFET sensor, one control and one readout ASIC.}
\end{figure}

For a source-drain current $I_\text{DS}$ of \SI{100}{\uA} $g_\text{q}$ is about \SI{700}{\pA} per collected electron.\cite{bonholzer22} The DEPFET layout for the WFI is shown in \cref{fig:depfet}. It features a linear gate layout which enables a better scalability than the more compact circular gate layout. Due to shorter gate widths the charge removal is improved significantly. Clear contacts on both ends of the gate act as drain for the electrons collected in the internal gate. The charge removal is switched on by appropriate voltages at the clear contacts and the so-called clear gate.

\section{DETECTOR OPERATION}

\begin{figure}[ht]
	\begin{center}
		\begin{tabular}{c}
			\fbox{\includegraphics[width=7cm]{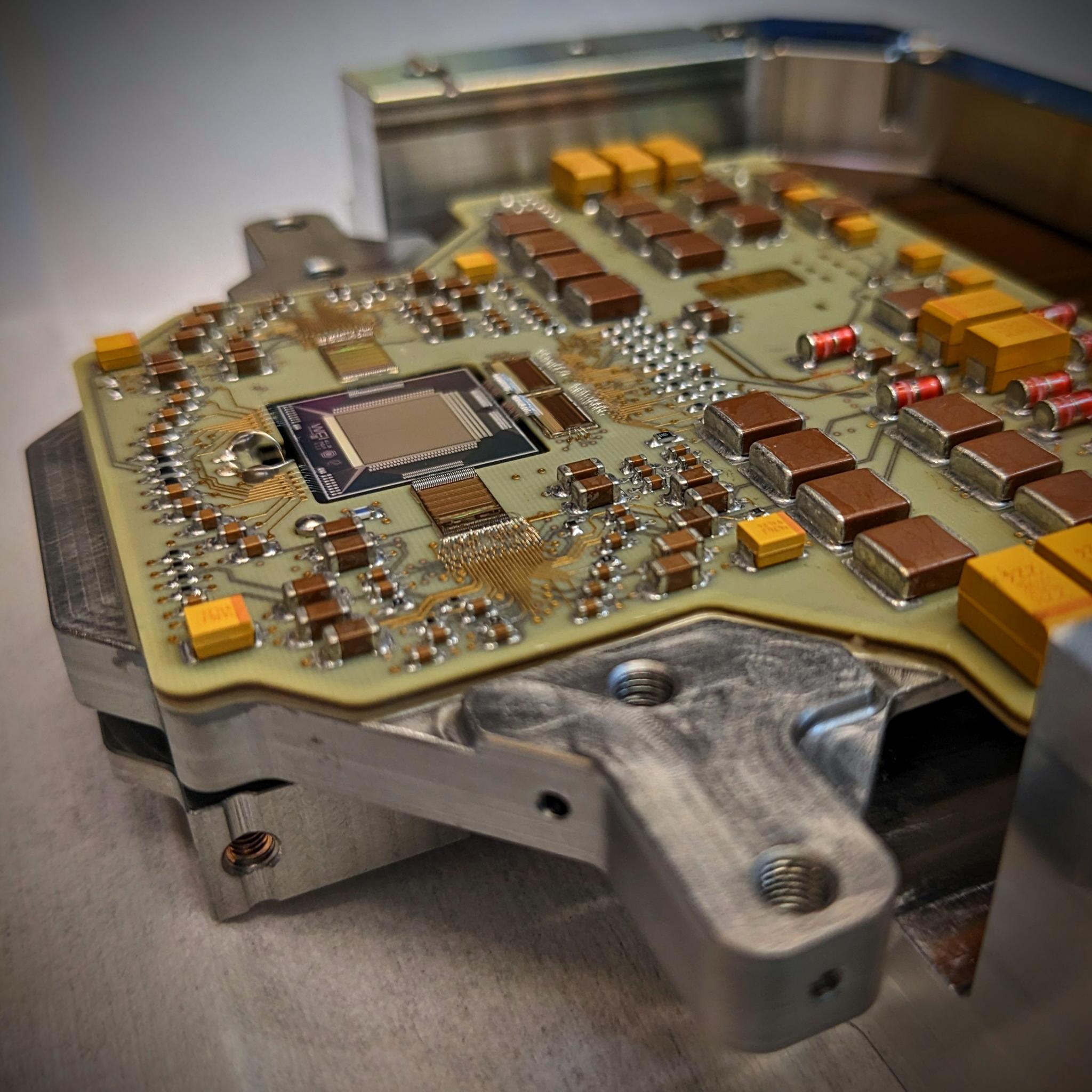}}\hspace{1.5cm}%
			\fbox{\includegraphics[width=7cm]{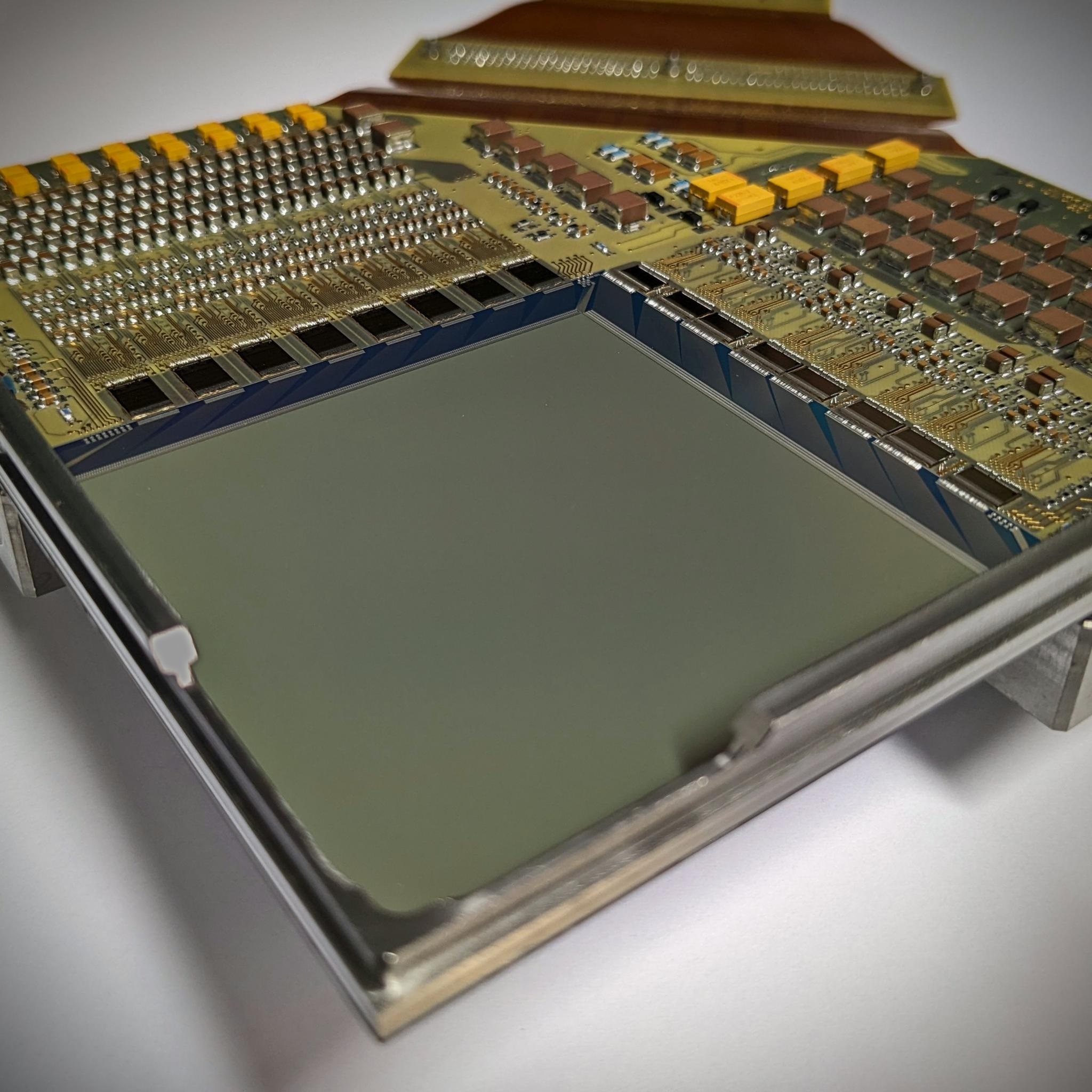}}%
		\end{tabular}
	\end{center}
	\caption[modules]
		{ \label{fig:modules} 
		A Fast Detector (left) and a Large Detector module (right). The front sides with the control and readout ASICs as well as the support electronics are shown. The photon entry sides of both detectors each point downwards and are therefore not visible. The complex cooling structures of the laboratory detector modules are according to the current state of the WFI mechanical design.}
\end{figure}

To operate detectors of various sizes and types, a flexible and modular laboratory system has been developed at the Max-Planck-Institute for Extraterrestrial Physics. It consists of a data acquisition system and power supplies for the detector electronics. The data acquisition needs to be fast enough to process all the pixel data taken by the detectors. Therefore, the laboratory system was developed based on the µTCA (Micro Telecommunications Computing Architecture) standard. The data acquisition system includes a sequencer in order to synchronise and command the control and readout electronics and the ADCs (Analogue-to-Digital Converter). This multichannel sequencer system consists of an FPGA (Field Programmable Gate Array) card and a sequencer interface. Analogue signals which are read out of the detectors are converted by two 14-bit ADC cards (four channels each) in combination with two FPGA cards. A Gigabit Ethernet interface with fibre optics per ADC channel performs the transmission of the frame data to the measurement computer. In order to minimise the data overhead, the transmission is based on UDP (User Datagram Protocol).

\subsection{Control ASIC}

The control electronics for the WFI DEPFETs is the Switcher ASIC (Application-Specific Integrated Circuit).\cite{fischer03} A version developed to enable the WFI detector configuration is able to switch three different signals at every of its 64 channels. They are used to switch the (external) DEPFET gates as well as the clear contacts and the clear gates of an entire sensor row simultaneously. Using the Switcher ASIC's internal shift register, consecutive channels can be selected one after another. Selected channels can be powered and activated to be ready to switch each of the three output voltages to control the DEPFET sensor's behaviour.

\subsection{Readout ASIC}

The Veritas (VErsatile Readout based on Integrated Trapezoidal Analog Shapers) is a mixed-signal readout ASIC realised in \SI{0.35}{\um} \SI{3.3}{\V} CMOS technology. The ASIC can read out 64 DEPFET pixels in parallel, generating a processed and serialised analogue stream at its output. It is operated in drain current readout mode, where the DEPFET drains of one detector row are connected to the inputs of the analogue channels of the Veritas. The DEPFET signals are amplified, shaped, and processed through two temporally sequenced measurements to obtain the signal step created by the removal of the accumulated charge due to X-ray photons. The processed signals are then multiplexed and buffered, resulting in a single serialised analogue output. This output is designed to connect to an ADC within the detector electronics for further digital processing. The output of data is performed while the next row of the sensor is already processed. Further, more detailed descriptions of the Veritas ASIC can be found in \ccite{porro14,herrmann18,schweingruber24}.

\begin{figure}[ht]
	\begin{center}
		\begin{tabular}{c}
			\includegraphics[width=13.2cm]{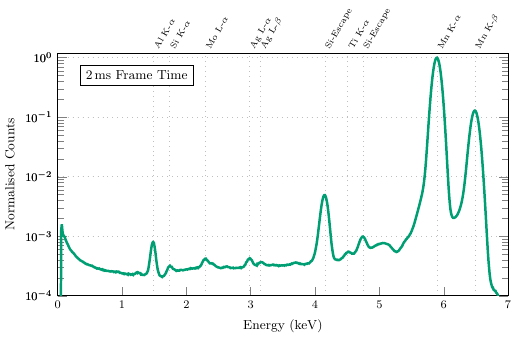}%
		\end{tabular}
	\end{center}
	\caption[spectrum]
		{ \label{fig:spectrum} 
		Iron-55 spectrum taken with a Large Detector, which was operated at a frame time of \SI{2}{\ms}. The energy resolution of the Mn~K-$\alpha$ emission line at \SI{5.9}{\keV} is \SI{131}{\eV}~FWHM.}
\end{figure}

The split readout of the Fast Detector requires two readout ASICs that process two rows at the same time and fully parallel. In principle, only one control ASIC is necessary to operate this $64\times64$ pixel sensor. For redundancy reasons one ASIC per detector half is used as visible in \cref{fig:modules}. The $512\times512$ pixels of a Large Detector require eight control and eight readout ASICs to operate it in the row-wise, column-parallel readout mode.

\section{MEASUREMENTS}
\label{sec:measu}

Electromagnetic radiation in the X-ray regime is usually characterised by its energy $E$. The conversion to frequency $\nu$ and wavelength $\lambda$ is given by the Planck relation.\cite{planck01}

\vspace{1.5em}
\begin{equation}
\label{eq:planck}
	\eqnmark[black]{erg}{E} = \eqnmark[red]{pc}{h} \eqnmark[blue]{fc}{\nu} = \frac{h \eqnmarkbox[yellow]{sol}{c}}{\eqnmark[green]{wl}{\lambda}}
\end{equation}
\annotate[yshift=-0.5em,xshift=-1em]{below,left}{erg}{Photon energy}
\annotate[yshift=1.5em]{above,left}{pc}{Planck constant}
\annotate[yshift=-2.5em,xshift=-1em]{below,left}{fc}{Photon frequency}
\annotate[yshift=0.7em]{above,right}{sol}{Speed of light}
\annotate[yshift=-1.7em,xshift=1em]{below,right}{wl}{Photon wavelength}
\vspace{3em}

Accordingly, the spectral performance is given as energy resolution. Two peaks can be separated from each other if the distance between them is bigger than the peak width at half amplitude. This limit width is called Full Width at Half Maximum (FWHM). For Fano-limited sensors like the WFI DEPFETs, the Fano noise\cite{fano47} is the dominating noise contribution. Further spectral degradation is caused by additional noise components in the sensor, the front-end electronics, which contains the ASICs, and the detector electronics that is needed to operate the cameras. The spectrum is degraded by broadening the Gaussian shape of an emission line but also by further components like threshold effects of events that split over more than one pixel. In this case, the line broadening deviates from a normal distribution. In addition, there are charge loss effects in the detector and degradation during the data processing. All these components are discussed in great detail in \ccite{mayr24}.

\begin{figure}[ht]
	\includegraphics[width=\textwidth]{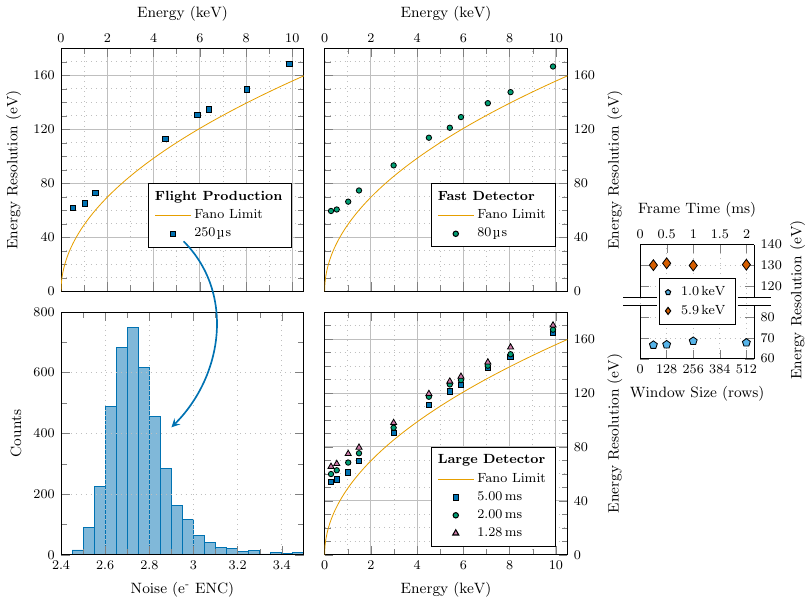}%
	\caption[perf]
		{ \label{fig:perf} 
		\textit{Upper left}: Spectral response of a prototype DEPFET detector from the flight production. The readout time per row equals the one of a Large Detector with a frame time of \SI{2}{\ms}. \textit{Upper centre}: Spectral response of a Fast Detector at its desired readout speed of \SI{80}{\us}. \textit{Lower left}: Noise characteristics of a prototype DEPFET detector from the flight production. \textit{Lower centre}: Spectral response of a Large Detector at different readout speeds. \textit{Right}: Spectral performance at \SI{1}{\keV} and \SI{5.9}{\keV} for different window sizes of a Large Detector.}
\vspace{1em}
\end{figure}

\vspace{2.5em}
\begin{equation}
\label{eq:fwhm}
	\eqnmark[black]{eres}{\Delta E} = \eqnmark[black]{fwhm}{\text{FWHM}(E)} = \eqnmark[black]{sig2fwhm}{2 \sqrt{2 \ln{2}}} {\color{green}\sqrt{\eqnmark[green]{fano}{F \omega E} + {\color{black}\eqnmark[black]{eherg}{\omega}^2 \eqnmarkbox[yellow]{noise}{\Sigma_j {\sigma_j}^2}}}} + \eqnmark[blue]{misc}{\Sigma_i \Delta \text{FWHM}_i(E)}
\end{equation}
\annotate[yshift=1.5em]{above,left}{eres}{Energy resolution}
\annotate[yshift=1.5em]{above,right}{fwhm}{Full width at half maximum at energy $E$}
\annotate[yshift=1.5em,xshift=-2em]{above,right}{misc}{Further line broadening effects}
\annotate[yshift=-1em,xshift=-1em]{below,left}{sig2fwhm}{Conversion factor from $\sigma$ to FWHM}
\annotate[yshift=-3em,xshift=-1em]{below,left}{fano}{Fano noise, $F \equiv$ Fano factor}
\annotate[yshift=-1em,xshift=1em]{below,right}{noise}{Further noise components $\sigma_j$}
\annotate[yshift=-3em,xshift=1em]{below,right}{eherg}{Mean electron-hole pair creation energy $\omega$}
\vspace{3.5em}

Most measurements are performed with an iron-55 source which emits predominantly photons at \SI{5.9}{\keV} (\cref{fig:spectrum}). To characterise the spectral response of the DEPFET sensors also at different energies within the targeted energy range from \SI{0.2}{\keV} to \SI{15}{\keV}, an X-ray tube is used. The continuous background generated by Bremsstrahlung requires additional effort in the data analysis which is described in \ccite{ms22,emberger22}. The measurement results with the Large and the Fast Detector were extended to higher energies and are shown in \cref{fig:perf}. Above \SI{8}{\keV} a significant fraction of photons is already absorbed near the front side of the sensor. This results in a performance degradation as known from the entrance window on the back side for low energetic photons below \SI{2}{\keV}.

\begin{figure}[ht]
	\begin{center}
		\begin{tabular}{c}
			\includegraphics[width=14.7cm]{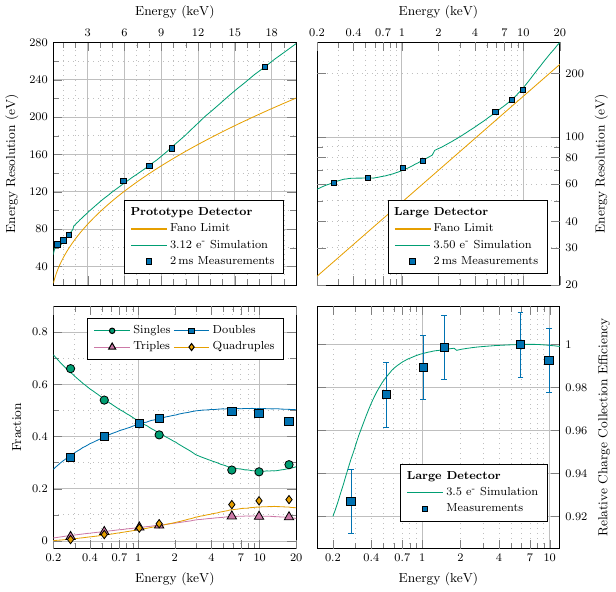}%
		\end{tabular}
	\end{center}
	\caption[calibration]
		{ \label{fig:calibration} 
		\textit{Upper left}: Measured spectral response including Mo~K-$\alpha$ at \SI{17.4}{\keV} to calibrate the charge losses at the front side. \textit{Upper right}: Measured spectral response with a focus on low energetic photons to calibrate charge losses at the photon entrance window (back side). \textit{Lower left}: Comparison of simulated (solid lines) and measured (marks) split fractions. \textit{Lower right}: Comparison of simulated and measured charge collections efficiencies. The data are normalised to \SI{5.9}{\keV}. The measurement errors are large because they base on the pixel-wise gain determination which is based on limited statistics.}
\end{figure}

The availability of first wafers from the flight production which was performed by the Semiconductor Laboratory of the Max-Planck-Society (HLL) also allowed for first tests with $64\times64$ prototype DEPFET sensors to verify the overall spectral performance quality of the fabrication. First electrical measurements by HLL already indicated an outstanding yield even for the $512\times512$ DEPFET sensors of \SI{>90}{\percent}. The spectroscopic measurements now also confirmed the excellent Fano-limited performance (see \cref{fig:perf}).

\section{SIMULATIONS}
\label{sec:simul}

\begin{figure}[ht]
	\begin{center}
		\begin{tabular}{c}
			\includegraphics[width=15cm]{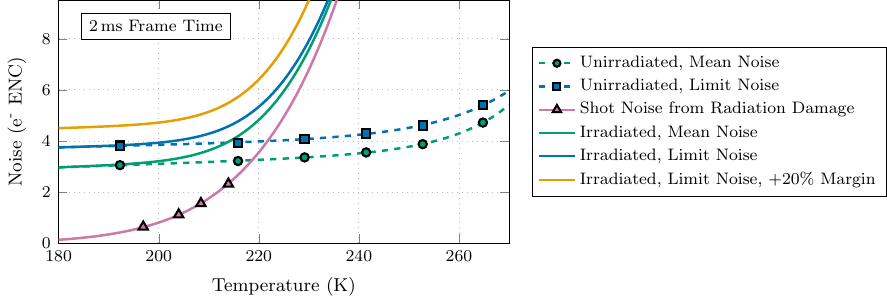}%
		\end{tabular}
	\end{center}
	\caption[noise]
		{ \label{fig:noise} 
		Fitted and calculated noise behaviour in dependence of the DEPFET sensor temperature. The proton irradiation campaign to determine the shot noise from radiation damage is presented in \ccite{emberger24}. Measured valued are represented by marks, while fits and calculations a represented by solid and dashed lines.}
\end{figure}

While measurements provide reliable results to describe the quality of freshly fabricated photo sensors, requirements for these radiation detectors for space missions are given for the end of life performance. After a few years in space, the performance degrades due to the bombardment of the instruments by high energetic particles. The effects of such a degradation by photons and massive particles can already be studied on ground by dedicated measurements as performed and described in \ccite{emberger22,emberger24}. Even though a slightly inhomogeneous illumination provides some knowledge of the radiation dose dependencies, usually only one or a few measurement points are available. On top, such measurement campaigns normally over-irradiate the devices to get more reliable statistics. Because not all components of \cref{eq:fwhm} can be derived analytically, the aim was to create a simple Monte Carlo simulation that is complex enough to describe the spectral response of the detectors. The purpose is to be able to take threshold effects into account of the performance assessment. Charge clouds can be distributed over multiple pixels. Small fractions of the charge cloud generated by an incident X-ray photon may be lost in the subsequent analysis because they are below the detection threshold. Even though these electrons were collected correctly in one of the neighbouring DEPFET pixels, they appear as apparent charge loss. This has a significant effect on the spectral resolution. The affect depends on the pixel size, the pixel-wise noise and the charge cloud size. While the first is known and the second value can be measured, the mean charge cloud size and its energy dependence was determined by the simulation itself.\cite{treberspurg18} The measured and the simulated split statistics were compared and iterated to find the mean charge cloud size $\sigma_\text{cloud}$ per energy. As a first simple approximation, the following relation is used.

\begin{equation}
\label{eq:ccs}
	\sigma_\text{cloud} = 
	\begin{dcases*}
		\left(\frac{E}{\SI{2745}{\eV}} + \frac{98}{9}\right)\unit{\um} & if $E < \SI{3050}{\eV}$ \\[1ex]
		\left(\frac{-E}{\SI{8269}{\eV}} + 12.68\right)\unit{\um} & if $E > \SI{5622.92}{\eV}$ \\[1ex]
		\SI{12}{\um} & otherwise
	\end{dcases*}
\end{equation}

Comparing measurements and simulation results as shown in \cref{fig:calibration} reveals, that the deviation gets larger for higher energies. This is likely due to the inhomogeneous electric field at the front side which cannot be approximated by mean values that are based on the photon energy. But because threshold effects play a minor role for high energetic particles and their large amounts of generated electron-hole pairs, this was not further investigated.

The second part that was calibrated by comparing measurements with simulations are the charge losses at the sensor surfaces. The exponentially distributed actual penetration depths of individual events are simulated based on X-ray attenuation lengths from \ccite{henke93}. Depending on the penetration depth $z$ in the sensor a fixed amount of charge was subtracted by multiplying with the charge collection efficiency $\eta_\text{coll}$ as a factor.

\begin{equation}
\label{eq:cce}
	\eta_\text{coll}(z) = 1 - 0.14\, e^{-z/\SI{0.045}{\um}} - 0.03\, e^{(z - \SI{450}{\um})/\SI{78}{\um}}
\end{equation}

\begin{figure}[ht]
	\begin{center}
		\begin{tabular}{c}
			\includegraphics[width=15cm]{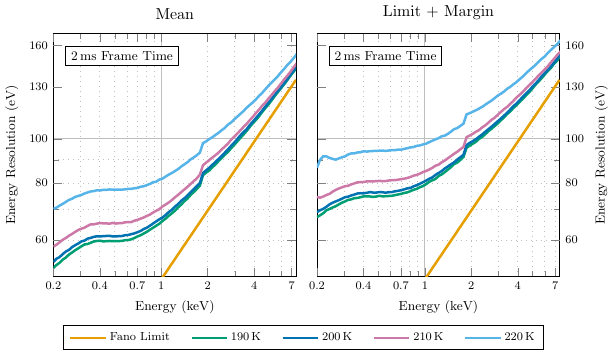}%
		\end{tabular}
	\end{center}
	\caption[simulation]
		{ \label{fig:simulation} 
		Simulated spectral responses for a detector with \SI{2}{\ms} frame time. On the left, the mean value over all pixels is simulated. On the right, the noise values of more noisy pixels plus a margin are used to show the limit of the achievable performance.}
\end{figure}

The values could be improved with more measurements at high photon energies. Like for the split statistics, the result for high energetic photons will also depend on the position on the sensor to account for the inhomogeneous sensor front side.

After these two effects were calibrated, the energy resolution delivered by the Monte Carlo simulation should depend only on the noise as input parameter. About at least one million events per energy are required to get reliable results. To cover the relevant region, 100 energies were distributed logarithmically over the range of \SI{0.2}{\keV} to \SI{20}{\keV}. The position in $x$, $y$ and $z$ as well as the noise were generated randomly, charge losses applied according to $z$ and the photon energy and the distribution of charge carriers on the central and neighbouring pixels was calculated applying the noise threshold. The collected data for an incident photon-energy were binned and the peak width was determined. Using the noise values taken from the data shown in \cref{fig:noise}, simulations for different energies were performed to determine the influence of radiation damage induced shot noise on the energy resolution.

\cref{fig:simulation} shows the mean energy resolution of a detector but also the performance for the worst pixels after a cut-off was applied allowing some pixels not to fulfil the performance requirements. The large margin of \SI{20}{\percent} should account for uncertainties arising from the limited data available about the degradation. Further measurement campaigns in the future will improve the prediction about the noise after years in space.

The input data are based on measurements with the laboratory setup only. The future usage of flight hardware might also decrease the noise performance.

\section{CONCLUSION}
\label{sec:concl}
Measurements with full-scale detectors, using the final layout and technology show excellent Fano-limited spectral performance. The DEPFET sensors were operated in a relevant environment such as a vacuum and appropriate, independent temperatures at the photon detector and the control and readout electronics. Radiation damage, which was investigated in separate campaigns, was taken into account by analysis and simulation to improve the prediction of end of life performance in space. Furthermore, such a simulation tool developed to analyse non-measurable quantities will aid in further optimising the instrument. First measurement results from the flight production confirm the quality of sensors fabricated at MPG's Semiconductor Laboratory. The fact that the effort to greatly improve the yield for large-area detectors was successful is also a reason to celebrate for the WFI instrument.

\acknowledgments
 
Development and production of the DEPFET sensors for the Athena WFI is performed in a collaboration between MPE and the MPG Semiconductor Laboratory (HLL). We gratefully thank all people who gave aid to make the presented results possible. The work was funded by the Max-Planck-Society and the German space agency DLR (FKZ: 50 QR 1901 and 50 QR 2301).

\bibliography{spie-13093-30_mueller-seidlitz} 

\begin{thebibliography}{10}

\bibitem{nandra13}
Nandra, K. et~al., ``{The Hot and Energetic Universe - A White paper presenting
  the science theme motivating the Athena+ mission},'' (2013).

\bibitem{barret23}
Barret, D., Albouys, V., Herder, J., et~al., ``{The Athena X-ray Integral Field
  Unit: a consolidated design for the system requirement review of the
  preliminary definition phase},'' {\em Exp Astron}~{\bf 55},  373--426 (2023).

\bibitem{henke93}
Henke, B.~L. et~al., ``{X-Ray Interactions: Photoabsorption, Scattering,
  Transmission, and Reflection at E = 50-30,000 eV, Z = 1-92},'' {\em Atomic
  Data and Nuclear Data Tables}~{\bf 54}(2),  181--342 (1993).

\bibitem{mms22}
Meidinger, N. and M{\"u}ller-Seidlitz, J.,  [{\em DEPFET Active Pixel
  Sensors}{\nolinebreak\hspace{0.1em}]},  1--20, Springer Nature Singapore,
  Singapore (2022).

\bibitem{kemmer87}
Kemmer, J. and Lutz, G., ``New detector concepts,'' {\em Nucl. Instr. Meth.
  Phys. Res. A}~{\bf 253},  365--377 (1987).

\bibitem{gatti84}
Gatti, E. and Rehak, P., ``Semiconductor drift chamber --- an application of a
  novel charge transport scheme,'' {\em Nucl. Instr. Meth. Phys. Res.}~{\bf
  225}(3),  608 -- 614 (1984).

\bibitem{fano47}
Fano, U., ``Ionization yield of radiations. ii. the fluctuations of the number
  of ions,'' {\em Phys. Rev.}~{\bf 72},  26--29 (1947).

\bibitem{lutz99}
Lutz, G.,  [{\em Semiconductor Radiation
  Detectors}{\nolinebreak\hspace{0.1em}]}, Springer, 1st~ed. (1999).

\bibitem{bonholzer22}
Bonholzer, M. et~al., ``{Drain current characteristics of Athena WFI
  flight-like DEPFETs},'' in [{\em Space Telescopes and Instrumentation 2022:
  Ultraviolet to Gamma Ray}{\nolinebreak\hspace{0.1em}]},   {\bf 12181}, Proc.
  of SPIE (2022).

\bibitem{fischer03}
Fischer, P. et~al., ``{Readout concepts for DEPFET pixel arrays},'' {\em Nucl.
  Instr. Meth. Phys. Res. A}~{\bf 512}(1),  318 -- 325 (2003).

\bibitem{porro14}
Porro, M. et~al., ``{VERITAS 2.0 a multi-channel readout ASIC suitable for the
  DEPFET arrays of the WFI for Athena},'' in [{\em Space Telescopes and
  Instrumentation 2014: Ultraviolet to Gamma Ray}{\nolinebreak\hspace{0.1em}]},
    {\bf 9144}, SPIE (2014).

\bibitem{herrmann18}
Herrmann, S. et~al., ``{VERITAS 2.2: a low noise source follower and drain
  current readout integrated circuit for the wide field imager on the Athena
  x-ray satellite},'' in [{\em High Energy, Optical, and Infrared Detectors for
  Astronomy VIII}{\nolinebreak\hspace{0.1em}]},   {\bf 10709},  741 -- 748,
  International Society for Optics and Photonics, SPIE (2018).

\bibitem{schweingruber24}
Schweingruber, A. et~al., ``{The VERITAS 2.3 readout ASIC for the ATHENA Wide
  Field Imager},'' in [{\em Space Telescopes and Instrumentation 2024:
  Ultraviolet to Gamma Ray}{\nolinebreak\hspace{0.1em}]},   {\bf 13093}, Proc.
  of SPIE (2024).

\bibitem{planck01}
Planck, M., ``{Ueber das Gesetz der Energieverteilung im Normalspectrum},''
  {\em Annalen der Physik}~{\bf 309}(3),  553--563 (1901).

\bibitem{mayr24}
Mayr, A. et~al., ``{Spectral Performance Budget for ATHENA’s Wide Field
  Imager},'' in [{\em Space Telescopes and Instrumentation 2024: Ultraviolet to
  Gamma Ray}{\nolinebreak\hspace{0.1em}]},   {\bf 13099}, Proc. of SPIE (2024).

\bibitem{ms22}
M{\"u}ller-Seidlitz, J. et~al., ``{Spectroscopic performance of flight-like
  DEPFET sensors for Athena's WFI},'' in [{\em Space Telescopes and
  Instrumentation 2022: Ultraviolet to Gamma Ray}{\nolinebreak\hspace{0.1em}]},
   den Herder, J.-W.~A., Nikzad, S., and Nakazawa, K., eds.,  {\bf 12181},
  121813W, International Society for Optics and Photonics, SPIE (2022).

\bibitem{emberger22}
Emberger, V. et~al., ``{Total ionizing dose test with DEPFET sensors for
  Athena's WFI},'' in [{\em Space Telescopes and Instrumentation 2022:
  Ultraviolet to Gamma Ray}{\nolinebreak\hspace{0.1em}]},   {\bf 12181}, Proc.
  of SPIE (2022).

\bibitem{emberger24}
Emberger, V. et~al., ``{Low temperature proton irradiation with DEPFETs for
  Athena’s WFI},'' in [{\em Space Telescopes and Instrumentation 2024:
  Ultraviolet to Gamma Ray}{\nolinebreak\hspace{0.1em}]},   {\bf 13093}, Proc.
  of SPIE (2024).

\bibitem{treberspurg18}
Treberspurg, W. et~al., ``{Energy response of ATHENA WFI prototype
  detectors},'' in [{\em Space Telescopes and Instrumentation 2018: Ultraviolet
  to Gamma Ray}{\nolinebreak\hspace{0.1em}]},   106994F, Proc. of SPIE (2018).

\end{thebibliography}
\bibliographystyle{spiebib} 

\end{document}